\documentclass[twocolumn]{aastex631}

\shorttitle{M71E Optical Identification}
\shortauthors{Liu \& Dong}

\graphicspath{{./}{figures/}}

\begin{document}

\title{Optical Identification of the Shortest-Period Spider Pulsar System M71E}

\correspondingauthor{Subo Dong}
\email{dongsubo@pku.edu.cn}

\author{Zhuokai Liu}
\affil{Department of Astronomy, School of Physics, Peking University, 
Yiheyuan Rd. 5, Haidian District, Beijing, China, 100871 \\}
\affil{Kavli Institute of Astronomy and Astrophysics, Peking University, 
Yiheyuan Rd. 5, Haidian District, Beijing, China, 100871 \\}
\author{Subo Dong}
\affil{Department of Astronomy, School of Physics, Peking University, 
Yiheyuan Rd. 5, Haidian District, Beijing, China, 100871 \\}
\affil{Kavli Institute of Astronomy and Astrophysics, Peking University, 
Yiheyuan Rd. 5, Haidian District, Beijing, China, 100871 \\}

\begin{abstract}
M71E is a spider pulsar (i.e., a millisecond pulsar with a tight binary companion) with the shortest known orbital period of $P=53.3$\,min discovered by Pan et al. (2023). Their favored evolutionary model suggests that it bridges between two types of spider pulsars, namely, it descended from a ``redback'' and will become a ``black widow''. Using {\it Hubble Space Telescope} (HST) archival imaging data, we report the first optical identification of its companion COM-M71E. The HST and pulsar timing coordinates are in excellent agreement (within $\sim10$\,mas). If M71E is associated with the globular cluster M71, our measured brightness of COM-M71E ($m_{\rm F606W} \approx 25.3$) is broadly consistent with the expectation from Pan et al. (2023)'s preferred binary evolutionary model of a stripped dwarf companion, while it is also compatible with an ultra-low-mass degenerate companion. Future multi-wavelength photometric and spectroscopic observations can characterize the companion and test the evolutionary scenarios.
\end{abstract}

\keywords{Millisecond pulsars (1062) ---  Binary pulsars (153) --- Optical identification (1167) --- Globular clusters (656)}

\section{Introduction} \label{sec:intro}
Millisecond pulsars (MSPs), rapidly-spinning pulsars (PSRs) found in either the Galactic field or globular clusters (GCs), are thought to be ``recycled'' old neutron stars, which have spun up via accreting mass from binary companions \citep{Alpar1982, BH1991}. In the classic picture, the accretion process occurs as low-mass X-ray binaries (LMXBs), after which MSPs are left with companions evolving into He white dwarfs (WDs). Most binary MSPs have WD companions following the mass-period relation expected from binary evolution models leading to He WDs \citep{HeWD1999}. However, the spider pulsars, with short orbital periods ($P\lesssim1$\,d) and primarily found in eclipsing systems, are exceptions to such a relation \citep[see, e.g.,][]{Roberts2013}: The ``black widows'' (BWs) have very low-mass companions ($M_{\rm c}\ll0.1\,M_\odot$), and in contrast, the companions of ``redbacks'' (RBs) are comparatively more massive ($M_{\rm c}\sim 0.1-0.4\,M_\odot$). The companion's evolution is affected by irradiation/ablation by the pulsar wind -- BWs may eventually become isolated MSPs after the evaporating companions are obliterated, and  \citet{Benvenuto2014} suggest that BWs descend from ablated RBs. The formation mechanisms of spider pulsars are debated \citep[e.g.,][]{King2003, King2005, Chen2013, Benvenuto2014, Jia2015, Ginzburg2020}. 

By analyzing the pulsar timing data from the Five-hundred-meter Aperture Spherical radio Telescope (FAST), \citet{pan_binary_2023} (hereafter Pan23) find that M71E (a.k.a., PSR J1953+1844, discovered by \citealt{Han2021} in the field of globular cluster M71) is an extraordinary non-eclipsing spider pulsar, which has a record-breakingly short orbital period of $P=53.3$\,min and a very small mass function of $2.3\times10^{-7}M_\odot$. Pan23's analysis favors a stripped dwarf companion with $M_c = 0.047 - 0.097\,M_\odot$, estimated by imposing mass-radius relations for brown dwarfs (BDs)/low-mass dwarfs \citep{Burrows1993} and restricting its radius within the Roche lobe radius \citep{Eggleton1983}. Such a moderate mass and its very short period suggest that it evolved from a RB and will evolve into a BW. Their favored binary evolution model based on \citet{Chen2013} expects that the companion has an effective temperature $T_{\rm eff}\sim4500$\,K and luminosity $L\sim3.25\times10^{-3}\,L_\odot$, which is much fainter than their upper limit on the optical brightness ($>22$\,mag) from non detections on SDSS images. 

In this paper, we detect the optical companion (hereafter COM-M71E) of M71E on an archival HST image. Our detection adds to a handful of optically identified GC spiders with very low-mass companions \citep[e.g.,][]{COM-M5, COM-M71}. 
\begin{figure*}[htp]
  \centering
  \includegraphics[width=0.5\textwidth]{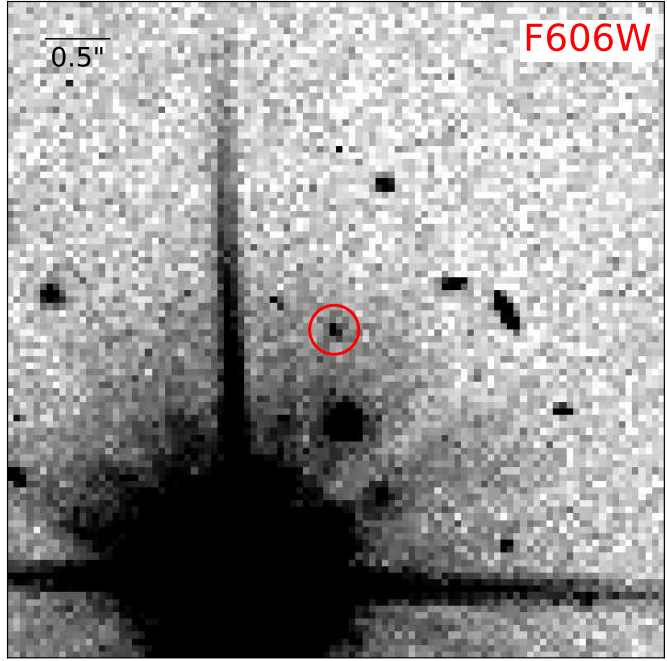}
  \caption{A $5\arcsec\times5\arcsec$ section of the HST ACS F606W image around COM-M71E. The red circle is centered at the pulsar timing position of M71E, which agrees with the HST F606W position within $\sim10$\,mas, and the circle's radius is 10 times of the HST astrometric error.}
  \label{fig:detection}
\end{figure*}

\section{Optical Observations \& Data Analysis} \label{sec:data}

We query the Mikulski Archive for Space Telescopes for archival HST images covering the location of M71E, which is $\sim 2.5\arcmin$ away from the center of M71. We find a single image (see Figure \ref{fig:detection} for a section centered on the location of M71E derived from pulsar timing) taken with the F606W filter and an exposure time of 339\,s by the Advanced Camera for Surveys (ACS) on UT November 22, 2021 (mid exposure time at MJD = 59540.11878) associated with program 16871\footnote{All the {\it HST} data used in this paper can be found in MAST: \dataset[10.17909/xb70-ee36]{http://dx.doi.org/10.17909/xb70-ee36}} (PI: J. Anderson; \citealt{2021hst..prop16871A}), which was proposed to make astrometric correction of the charge transfer efficiency (CTE) effects. Visual inspections near the M71E position reveal a faint point source that is $\sim 2\arcsec$\,away from a bright star. 

We employ the software \texttt{hst1pass}\footnote{\url{https://www.stsci.edu/~jayander/HST1PASS/}} \citep{2022AAS...24020601A} based on the effective point-spread function (ePSF) method \citep{ePSF} to perform astrometric and photometric analysis using the pixel-by-pixel CTE-corrected data product (i.e., the image with ``\_flc'' suffix).  We set the parameters ${\rm FMIN}$ = 100 and $\rm HMIN = 3$ to allow detecting faint sources and finding stars near bright neighbors, respectively. The faint source close to M71E visible to by-eye inspections is detected by \texttt{hst1pass}. To investigate possible issues of extracting a faint source in the proximity of a bright star, we inject 10 artificial stars with the same instrumental magnitude as the source of interest in the vicinity of a bright star sharing a similar background level and gradient, and the injections are at random sub-pixel positions. We recover the astrometric position within $\sim5$\,mas and magnitude within $\sim0.1$\,mag, suggesting that the source extraction is reliable.   

We subsequently compare the resulting star catalog with the HST ACS F606W results of M71 taken on UT May 12, 2006 in \citet{2007AJ....133.1658S}. We perform zero-point magnitude calibration using the common stars, and we determine that COM-M71E has $m_{\rm F606W} = 25.34\pm0.31$ in the VEGAmag system. We also estimate the relative astrometric uncertainties as a function of magnitude from epoch-to-epoch comparisons. Then we perform astrometric calibration using common stars with the Gaia Data Release 3 (DR3) catalog \citep{Gaia2016, GaiaDR3}. We apply 2nd-order 2D-polynomial transformations of the distortion-corrected ``(r,d)'' coordinates from \texttt{hst1pass} to the Gaia DR3 frame and translate the Gaia coordinates to the HST epoch by considering the proper motions. Our derived J2000.0 position of COM-M71E is (RA, Dec) $=(19^{\rm h}53^{\rm m}37.947^{\rm s}\pm0.001^{\rm s}, 18\arcdeg44\arcmin54.32\arcsec \pm 0.01\arcsec)$, and the uncertainties are estimated by combining the relative error derived from HST epoch-to-epoch comparisons and the Gaia calibration error. The comparison of the HST position with Pan23's pulsar timing position at (RA, Dec) $= (19^{\rm h}53^{\rm m}37.9464^{\rm s}\pm0.0001^{\rm s}, 18\arcdeg44\arcmin54.310\arcsec\pm0.002\arcsec)$ shows an excellent agreement (within $\sim10$\,mas) at the $<1\,\sigma$ level. Therefore, the HST detection of COM-M71E is a secure optical identification.

We assume that M71E is associated with M71, which has a distance of $D = 4.0$\,kpc \citep{GCdist}. We adopt an extinction correction of $A_{\rm F606W} = 0.6$ estimated using the M71 reddening value from \citet{Dotter2010} and the extinction coefficient from \citet{2005PASP..117.1049S}. Then we obtain COM-M71's absolute magnitude $M_{\rm F606W} = 11.7\pm0.3$.

\begin{figure}[ht]
\includegraphics[width=0.5\textwidth]{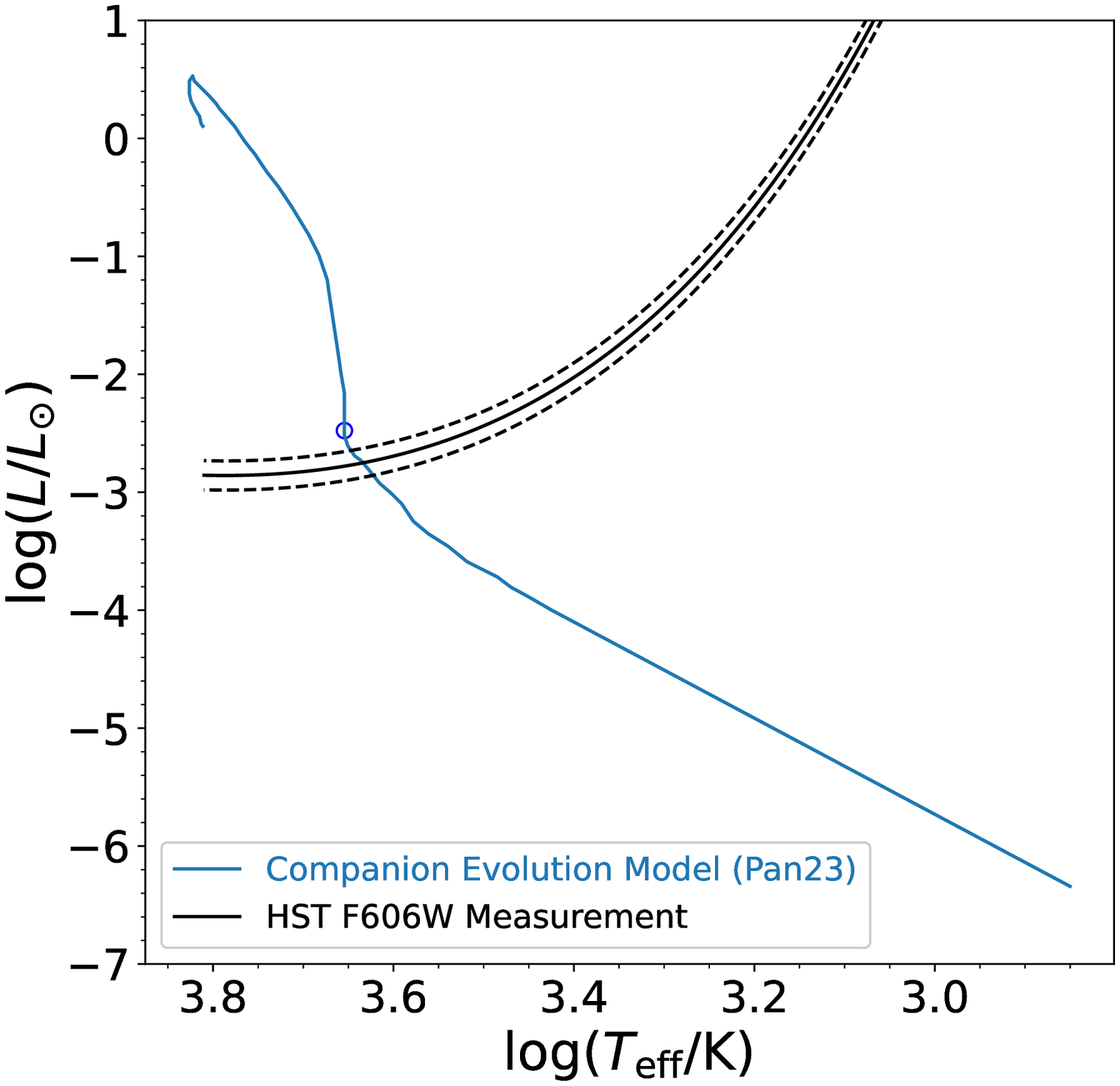}
\caption{Comparison between Pan23's favored binary evolutionary model and the observational constraints from HST photometry of COM-M71E in the Hertzsprung-Russell (HR) diagram. The blue curve is the evolutionary track presented in Panel (A) of Extended Data Figure 4 of Pan23. The blue open circle shows the estimated properties ($T_{\rm eff}\sim4500$\,K and luminosity $L\sim3.25\times10^{-3}\,L_\odot$) of the companion by Pan23. The solid black line shows the $T_{\rm eff}$ and $L$ constraint from the $M_{\rm F606W}$ derived from the HST optical detection, and the black dashed lines correspond to the 1$\sigma$  range. \label{fig:L_T}}
\end{figure}

\section{Discussion \& Summary} \label{sec:discussion}

With the detection only in a single band (F606W), it is not possible to independently derive COM-M71E's physical properties from optical. Nevertheless, we first make a comparison with the expected properties ($T_{\rm eff}\sim4500$\,K and $L\sim3.25\times10^{-3}\,L_\odot$) from Pan23's preferred binary-evolution model. We use \texttt{SYNPHOT} function of \texttt{MAAT} \citep{Ofek2014} to estimate the bolometric correction in F606W for a blackbody at $T_{\rm eff} = 4500$\,K, and find that the expected absolute magnitude is $M \sim 11.1$. Given the potential theoretical uncertainties, it is broadly consistent with $M_{\rm F606W} = 11.7\pm0.3$ from observation, supporting Pan23's interpretation that M71E is at an intermediate evolutionary stage between a RB and a BW. Under the assumption of this model, we can estimate the physical parameters of COM-M71E using our optical detection. By combining the theoretical $\log(T_{\rm eff})-\log(L)$ track of the companion (blue solid line in Figure \ref{fig:L_T}) and the F606W photometric constraints (black lines n Figure \ref{fig:L_T}), we obtain a blackbody radius $R_{\rm c, BB} \sim 0.08\,R_\sun$. Following Pan23, we subsequently estimate its mass $M_{\rm c} \sim 0.09\,M_\odot$ based on the mass-radius relations of \citet{Burrows1993}, and thus $R_{\rm c, BB}$ is smaller than the Roche lobe radius $\sim 0.1\,R_\odot$. We note some caveats regarding this interpretation. First of all, in Pan23's simulation, the BD donor's surface is He-rich, whereas \citet{Burrows1993}'s BD models assume solar abundance. Second, in this scenario, the companion was initially H-rich. While its orbital period ($P\approx53$\,min) is larger than the theoretically allowed minimum period of $P_{\rm min} \approx 37$\,min for a generic H-rich BD \citep{Rappaport2021}, it is smaller than the minimum value $P_{\rm min}$ either predicted by binary evolution $(P_{\rm min, theory}\sim 70$\,min; \citealt{KolbBaraffe1999}) or empirically seen in cataclysmic variables $(P_{\rm min, observed} \sim 80$\,min; \citealt{Knigge2006, Gansicke2009}) for mass transfer from such a donor. Lastly, reaching a mass of $M_{\rm c} \sim 0.09\,M_\odot$ requires an orbital inclination of $\lesssim 5$\,deg, which has a small geometric probability of $<1\%$ for randomly distributed orbital orientations.

Pan23 disfavor WD companions because a plausible WD with the minimum mass of $\sim0.16\,M_\odot$ would need a finely-tuned face-on orbit with a very small geometric probability ($<0.3\%$), and they also argue such a system is challenging to form from theory. Such a minimum mass is based on the observed low-mass WD population by \citet{Brown2016}. A WD with $\approx 0.16\,M_\odot$ needs to have a temperature $\approx10^4$K (typical for WDs in \citealt{Brown2016}) to match the observed F606W flux. A minimum WD mass of $\sim0.16\,M_\odot$ is also consistent with the classic evolution pathway by \citet{HeWD1999}, but it corresponds to a minimum period $\sim1$\,d, much larger than the observed $P\approx53$\,min. Alternatively, it may be possible to reach such a short period from a partially evolved donor star at the onset of mass transfer, and the orbital period shrinks during the mass transfer, eventually leading to mass loss from the WD core \citep[see, e.g.][]{Nelson1986, DeloyeBildsten2003}. Such a scenario may explain the extremely low-mass He/C-rich degenerate companions found in some MSP systems (e.g., \citealt{Bailes2011, Romani2012}). We briefly discuss the possibility that M71E is an ultra-low-mass He-rich WD. We examine a fiducial He WD with mass of $M_{\rm WD} = 0.02\,M_\odot$ corresponding to {\it a prior} likely inclination of $i\approx30^\circ$, with a radius of $R_{\rm WD} \approx 0.04\,R_\odot$ at $T_{\rm WD} \approx 6000$\,K according to \citet{DeloyeBildsten2003}, it can match the observed F606W flux.
Multi-band photometric measurements would permit a temperature estimate and could thereby test the WD scenario. Furthermore, spectroscopic observation can constrain its chemical composition (H/He/C) and provide further clues to its formation pathway.

Spider pulsar companions can be heated by PSR irradiation and can also be significantly tidally distorted, inducing flux changes over its orbital period. Known optical counterparts of spider pulsars can show significant photometric variations modulated by the orbital period, usually with maxima at inferior conjunctions (orbital phase $\phi = 0.75$) and minima at superior conjunctions ($\phi = 0.25$).  For example, the optical light curves of COM-M71A (a BW companion in M71) show periodic variations with a full amplitude $\sim 3$\,mag \citep{COM-M71}. Using Pan23's pulsar timing parameters (Time of Ascending Node at MJD $= 58829.26006$ in solar system's barycentric reference frame and orbital period $P = 0.0370398638$\,d), our HST observation (adding a -2.7\,min correction to transform into the same reference frame as pulsar timing) is at $\phi = 0.67$, which is close to the expected maximum. The light-curve amplitude is a function of orbital inclination \citep{COM-M5}, and if the orbit is nearly face-on as expected by P23, the M71E system would have much smaller variations than the eclipsing systems. 

In conclusion, the HST optical observation of COM-M71A broadly agree with the stripped dwarf model favored by Pan23. Future HST multi-band photometric monitoring and JWST spectroscopic observations will be desirable for testing various scenarios and making physical characterizations.
 \section*{Acknowledgment}
We are grateful to the anonymous reviewer for insightful comments and suggestions on the theoretical interpretation. We thank Jay Anderson, Hailiang Chen, Andy Gould, Kejia Lee, Zhichen Pan and Bing Zhang for helpful discussions. This work is supported by the science research grants from the China Manned Space Project with No. CMS-CSST-2021-B12, the National Natural Science Foundation of China (Grant No. 12133005) and the New Cornerstone Science Foundation through the XPLORER PRIZE. This research is based on observations made with the NASA/ESA Hubble Space Telescope obtained from the Space Telescope Science Institute, which is operated by the Association of Universities for Research in Astronomy, Inc., under NASA contract NAS 5-26555. These observations are associated with program 16871.  This work has made use of data from the European Space Agency (ESA) mission {\it Gaia} (\url{https://www.cosmos.esa.int/gaia}), processed by the {\it Gaia} Data Processing and Analysis Consortium (DPAC,
\url{https://www.cosmos.esa.int/web/gaia/dpac/consortium}). Funding for the DPAC has been provided by national institutions, in particular the institutions participating in the {\it Gaia} Multilateral Agreement. This research has made use of the VizieR catalogue access tool, CDS, Strasbourg, France.

\software{hst1pass, MAAT} 

\bibliography{ms}{}
\bibliographystyle{aasjournal}



\end{document}